\documentclass[twocolumn]{aastex631}
\usepackage{mathrsfs }
\usepackage{natbib}
\usepackage{amssymb}
\usepackage{apjfonts}
\usepackage{ulem} %Makes striking out text possible with \sout{}
\usepackage{graphicx}
\newcommand{\secpoint}{\mbox{$''\mskip-7.6mu.\,$}}

\begin{document}
\title{The AURORA Survey: An Extraordinarily Mature, Star-forming Galaxy at $z\sim 7$}

%\correspondingauthor{Alice Shapley}
%\email{aes@astro.ucla.edu}
\author[0000-0003-3509-4855]{Alice E. Shapley}\affiliation{Department of Physics \& Astronomy, University of California, Los Angeles, 430 Portola Plaza, Los Angeles, CA 90095, USA}
\email{aes@astro.ucla.edu}
%\maketitle

\author[0000-0003-4792-9119]{Ryan L. Sanders}\affiliation{Department of Physics and Astronomy, University of Kentucky, 505 Rose Street, Lexington, KY 40506, USA}

\author[0000-0001-8426-1141]{Michael W. Topping}\affiliation{Steward Observatory, University of Arizona, 933 N Cherry Avenue, Tucson, AZ 85721, USA}

\author[0000-0001-9687-4973]{Naveen A. Reddy}\affiliation{Department of Physics \& Astronomy, University of California, Riverside, 900 University Avenue, Riverside, CA 92521, USA}

\author[0000-0003-4464-4505]{Anthony J. Pahl}\affiliation{The Observatories of the Carnegie Institution for Science, 813 Santa Barbara Street, Pasadena, CA 91101, USA}

\author[0000-0001-5851-6649]{Pascal A. Oesch}\affiliation{Department of Astronomy, University of Geneva, Chemin Pegasi 51, 1290 Versoix, Switzerland}\affiliation{Niels Bohr Institute, University of Copenhagen, Lyngbyvej 2, DK2100 Copenhagen \O, Denmark}
\affiliation{Cosmic Dawn Center (DAWN), Copenhagen, Denmark}

\author[0000-0002-4153-053X]{Danielle A. Berg}\affiliation{Department of Astronomy, The University of Texas at Austin, 2515 Speedway, Stop C1400, Austin, TX 78712, USA}

\author[0000-0002-4989-2471]{Rychard J. Bouwens}\affiliation{Leiden Observatory, Leiden University, NL-2300 RA Leiden, Netherlands}

\author[0000-0003-2680-005X]{Gabriel Brammer}\affiliation{Niels Bohr Institute, University of Copenhagen, Lyngbyvej 2, DK2100 Copenhagen \O, Denmark}
\affiliation{Cosmic Dawn Center (DAWN), Copenhagen, Denmark}

\author[0000-0002-1482-5818]{Adam C. Carnall}\affiliation{Institute for Astronomy, University of Edinburgh, Royal Observatory, Edinburgh, EH9 3HJ, UK}

\author[0000-0002-3736-476X]{Fergus Cullen}\affiliation{Institute for Astronomy, University of Edinburgh, Royal Observatory, Edinburgh, EH9 3HJ, UK}

\author[0000-0003-2842-9434]{Romeel Dav\'e}\affiliation{Institute for Astronomy, University of Edinburgh, Royal Observatory, Edinburgh, EH9 3HJ, UK}

\author{James S. Dunlop}\affiliation{Institute for Astronomy, University of Edinburgh, Royal Observatory, Edinburgh, EH9 3HJ, UK}

\author[0000-0001-7782-7071]{Richard S. Ellis}\affiliation{Department of Physics \& Astronomy, University College London. Gower St., London WC1E 6BT, UK}

\author[0000-0003-4264-3381]{N. M. F\"orster Schreiber}\affiliation{Max-Planck-Institut f\"ur extraterrestrische Physik (MPE), Giessenbachstr.1, D-85748 Garching, Germany}

\author[0000-0002-0658-1243]{Steven R. Furlanetto}\affiliation{Department of Physics \& Astronomy, University of California, Los Angeles, 430 Portola Plaza, Los Angeles, CA 90095, USA}

\author[0000-0002-3254-9044]{Karl Glazebrook}\affiliation{Centre for Astrophysics and Supercomputing, Swinburne University of Technology, P.O. Box 218, Hawthorn, VIC 3122, Australia}

\author[0000-0002-8096-2837]{Garth D. Illingworth}\affiliation{Department of Astronomy and Astrophysics, UCO/Lick Observatory, University of California, Santa Cruz, CA 95064, USA}

\author[0000-0001-5860-3419]{Tucker Jones}\affiliation{Department of Physics and Astronomy, University of California Davis, 1 Shields Avenue, Davis, CA 95616, USA}

\author[0000-0002-7613-9872]{Mariska Kriek}\affiliation{Leiden Observatory, Leiden University, NL-2300 RA Leiden, Netherlands}

\author[0000-0003-4368-3326]{Derek J. McLeod}\affiliation{Institute for Astronomy, University of Edinburgh, Royal Observatory, Edinburgh, EH9 3HJ, UK}

\author{Ross J. McLure}\affiliation{Institute for Astronomy, University of Edinburgh, Royal Observatory, Edinburgh, EH9 3HJ, UK}

\author[0000-0002-7064-4309]{Desika Narayanan}\affiliation{Department of Astronomy, University of Florida, 211 Bryant Space Sciences Center, Gainesville, FL, USA}

\author{Max Pettini}\affiliation{Institute of Astronomy, Madingley Road, Cambridge CB3 OHA, UK}

\author[0000-0001-7144-7182]{Daniel Schaerer}\affiliation{Department of Astronomy, University of Geneva, Chemin Pegasi 51, 1290 Versoix, Switzerland}

\author{Daniel P. Stark}\affiliation{Steward Observatory, University of Arizona, 933 N Cherry Avenue, Tucson, AZ 85721, USA}

\author[0000-0002-4834-7260]{Charles C. Steidel}\affiliation{Cahill Center for Astronomy and Astrophysics, California Institute of Technology, MS 249-17, Pasadena, CA 91125, USA}

\author[0000-0001-5940-338X]{Mengtao Tang}\affiliation{Steward Observatory, University of Arizona, 933 N Cherry Avenue, Tucson, AZ 85721, USA}

\author[0000-0003-1249-6392]{Leonardo Clarke}\affiliation{Department of Physics \& Astronomy, University of California, Los Angeles, 430 Portola Plaza, Los Angeles, CA 90095, USA}

\author[0000-0002-7622-0208]{Callum T. Donnan}\affiliation{Institute for Astronomy, University of Edinburgh, Royal Observatory, Edinburgh, EH9 3HJ, UK}

\author{Emily Kehoe}\affiliation{Department of Physics \& Astronomy, University of California, Los Angeles, 430 Portola Plaza, Los Angeles, CA 90095, USA}

\shortauthors{Shapley et al.}

\shorttitle{An Extraordinary Evolved $z\sim 7$ Galaxy}

\begin{abstract}
We present the properties of a massive, large, dusty, metal-rich, star-forming galaxy at  $z_{\rm spec}=6.73$.  GOODSN-100182 was observed with {\it JWST}/NIRSpec as part of the AURORA survey, and is also covered by public multi-wavelength {\it HST} and {\it JWST} imaging. While the large stellar mass of GOODSN-100182 ($\sim10^{10} M_{\odot}$) was indicated prior to {\it JWST}, NIRCam rest-optical imaging now reveals the presence of an extended disk ($r_{\rm eff}\sim 1.5$~kpc).  In addition, the NIRSpec $R\sim1000$ spectrum of GOODSN-100182 includes the detection of a large suite of rest-optical nebular emission lines ranging in wavelength from [OII]$\lambda3727$ up to [NII]$\lambda6583$. The ratios of Balmer lines suggest significant dust attenuation ($E(B-V)_{\rm gas}=0.40^{+0.10}_{-0.09}$), consistent with the red rest-UV slope inferred for GOODSN-100182 ($\beta=-0.50 \pm 0.09$). The star-formation rate based on dust-corrected H$\alpha$ emission is $\log({\rm SFR(H}\alpha)/ {\rm M}_{\odot}{\rm yr}^{-1})=2.02^{+0.13}_{-0.14}$, well above the $z\sim7$ star-forming main sequence in terms of specific SFR. Strikingly, the ratio of [NII]$\lambda 6583$/H$\alpha$ emission suggests almost solar metallicity, as does the ratio ([OIII]$\lambda5007$/H$\beta$)/([NII]$\lambda 6583$/H$\alpha$) and the detection of the faint [FeII]$\lambda4360$ emission feature. Overall, the excitation and ionization properties of GOODSN-100182 more closely resemble those of typical star-forming galaxies at $z\sim2-3$ rather than $z\sim7$. Based on public spectroscopy of the GOODS-N field, we find that GOODSN-100182 resides within a significant galaxy overdensity, and is accompanied by a spectroscopically-confirmed neighbor galaxy. GOODSN-100182 demonstrates the existence of mature, chemically-enriched galaxies within the first billion years of cosmic time, whose properties must be explained by galaxy formation models. 
\end{abstract}

\section{Introduction}
Prior to the launch of {\it JWST}, our knowledge of the $z\sim 7$ galaxy population was limited almost exclusively to galaxy photometric candidates. From such samples, and observational datasets including ground-based, {\it Hubble Space Telescope} ({\it HST}), and {\it Spitzer}/IRAC imaging, it was possible to determine the evolving UV luminosity and stellar mass functions \citep{bouwens2015,stefanon2021}, the distribution of blue rest-frame UV colors \citep{wilkins2011,finkelstein2012,dunlop2013,bouwens2014},  and even (indirectly) the range of strong [OIII]+Hb emission-line equivalent widths \citep{endsley2023}.

{\it JWST} has opened a new window into the early galaxy population, enabling for the first time the direct spectroscopic characterization of the stars and gas in large samples of $z\sim 7$ galaxies \citep[e.g.,][]{curti2023,robertsborsani2024,meyer2024}. On average, the new spectroscopic observations of $z\sim 7$ galaxies reveal a star-forming, ionized ISM with significantly subsolar metallicities, high excitation, and minimal dust content \citep[e.g.,][]{sanders2023b,cameron2023,robertsborsani2024}. Furthermore, at fixed stellar mass, $z\sim 7$ galaxies have lower metallicity than their lower-redshift counterparts, reflecting the gradual evolution in the gas-phase mass-metallicity relationship \citep[MZR;][]{nakajima2023}. Finally, there is intriguing evidence that the balance that seems to hold among stellar mass, metallicity, and star-formation rate (SFR) from $z\sim 0$ back to $z\sim 3$, i.e.,  the Fundamental Metallicity Relation (FMR), may in fact evolve at $z>6$. Such $z>6$ galaxies show a deficit in metallicity relative to the lower-redshift FMR \citep{curti2024}. At the same time, {\it JWST} spectroscopy of H$\alpha$ emission has also uncovered a population of moderate-luminosity active galactic nuclei (AGN) at $z=4-7$, in which the ratio of black hole to stellar mass significantly exceeds the local relation \citep[e.g.,][]{maiolino2024,pacucci2023}, and the low ratio of [NII]$\lambda6583$/H$\alpha$ suggests subsolar metallicity.

While {\it JWST} has traced the relatively pristine properties of the bulk of the $z\sim 7$ star-forming galaxy population at $M_*<10^9 M_{\odot}$, its infrared imaging capabilities have also been used to identify a rare tail of massive and/or dusty galaxies at the same redshift. Such galaxies have significantly redder rest-UV/optical colors and many were missed in the previous {\it HST}-based census of $z>3$ galaxies \citep{labbe2023,barrufet2023,williams2024,gottumukkala2024,chworowsky2024}. Furthermore, some of these red galaxies, known as ``Little Red Dots," are characterized by extremely compact morphologies and appear to be dominated by AGN emission \citep{greene2024,kocevski2024}. Clearly, a full description of the $z\sim 7$ galaxy population needs to account for both its low-mass, blue, low-metallicity and high-mass, dusty, high-metallicity extremes. Here we describe the serendipitous discovery of a high-mass, large, dusty, high-metallicity star-forming galaxy at $z=6.73$ in the course of the Assembly of Ultradeep Rest-optical Observations Revealing Astrophysics (AURORA) Cycle~1 {\it JWST}/NIRSpec program \citep{shapley2024}.

The AURORA program was designed to detect faint auroral emission lines in $z>1.4$ galaxies and determine their direct metallicities. The long exposures required for such detections are also ideal for probing the spectroscopic properties of $z>6$ galaxies, as is the continuous wavelength coverage from $1-5 \mu{\rm m}$ using each of the three medium-resolution ($R\sim1000$) NIRSpec gratings. The galaxy GOODSN-100182 was targeted in AURORA as a $z>6$ candidate, simply on the basis of its photometric redshift. As described in this work, the high signal-to-noice (S/N) rest-optical spectrum of GOODSN-100182, along with its structural properties, place this galaxy in a truly unique part of physical parameter space. In particular, the mature chemical enrichment properties of GOODSN-100182 are atypical of this early cosmic epoch, and must be explained by any complete model of galaxy formation.

In \S\ref{sec:obs}, we describe our observations and data analysis. In \S\ref{sec:results}, we present several key physical properties of GOODSN-100182, based on both {\it JWST}/NIRCam imaging and NIRSpec spectroscopy. Finally, in \S\ref{sec:discussion}, we consider this remarkable galaxy in a larger context and discuss future directions, and summarize our key conclusions in \S\ref{sec:summary}.
Throughout this work, we adopt cosmological parameters of $H_0=70\mbox{ km  s}^{-1}\mbox{ Mpc}^{-1}$, $\Omega_{\rm m}=0.30$, and
$\Omega_{\Lambda}=0.7$, and a \citet{chabrier2003} IMF.

\begin{figure*}[t!]
\centering
\includegraphics[width=1.0\linewidth]{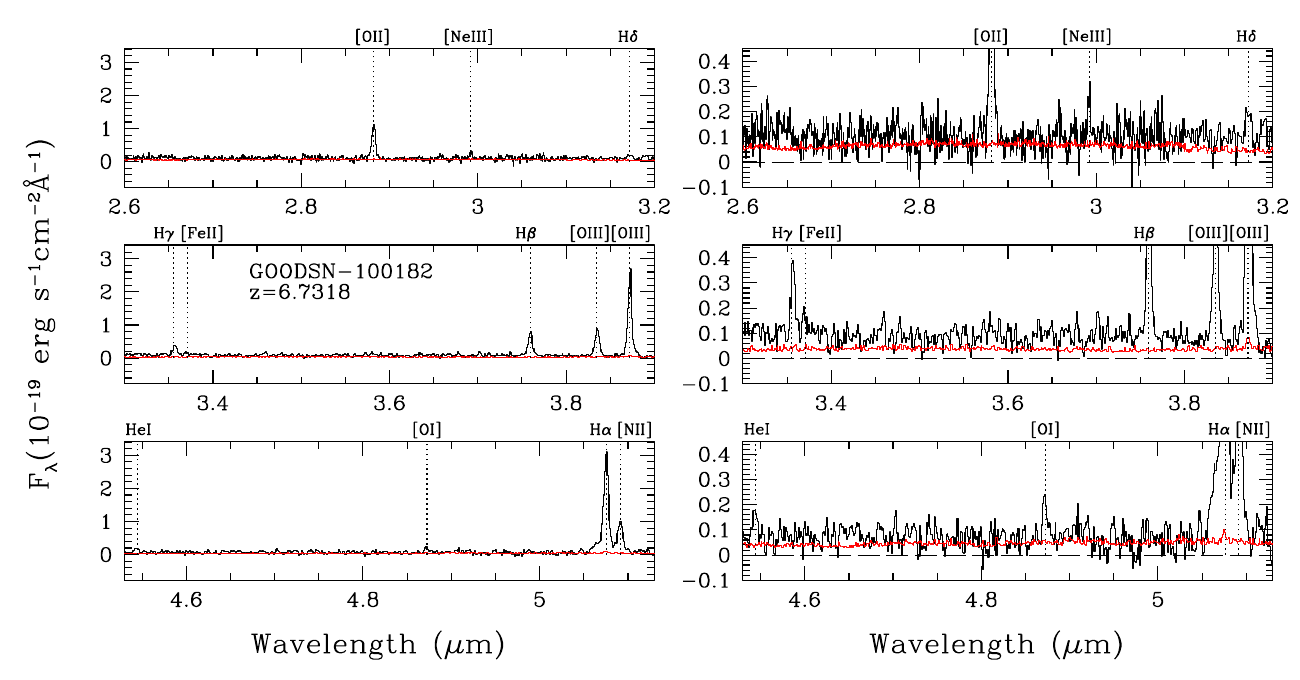}
\caption{NIRSpec rest-optical spectrum of GOODSN-100182. Flux density is plotted in black, while the error spectrum is plotted in red. The strongest rest-optical emission lines are labeled, spanning in wavelength from [OII]$\lambda 3727$ to [NII]$\lambda6583$, and their fluxes are reported in Table~\ref{tab:emline}. {\bf Left:} The spectrum with vertical scaling set by the strongest emission line (H$\alpha$). {\bf Right:} The spectrum covering the same wavelength ranges but with zoomed-in vertical scaling in order to showcase the faintest detected emission lines (e.g., H$\delta$, [FeII]$\lambda4360$, HeI$\lambda5877$, and [OI]$\lambda6300$). In the top panel, the transition between G235M and G395M gratings is evident at $\lambda=3.1\mu$m. At longer wavelengths than this transition, the spectrum noise level is decreased. Both middle and bottom panels feature G395M spectra.
}
\label{fig:spec-100182}
\end{figure*}

\section{Observations}
\label{sec:obs}
GOODSN-100182 (R.A.= 12:36:42.5184 and decl.= +62:17:29.462 (J2000)) was observed as part of the Assembly of Ultradeep Rest-optical Observations Revealing Astrophysics (AURORA) Cycle~1 {\it JWST}/NIRSpec program. As described in more detail in \citet{shapley2024}, AURORA consists of {\it JWST}/NIRSpec Micro Shutter Assembly  (MSA) observations in two separate pointings (one in the COSMOS field and one in GOODSN). Continuous wavelength coverage was obtained from 1--5$\mu$m at $R\sim 1000$ using the grating/filter combination of G140M/F100LP, G235M/F170LP, and G395M/F290LP. Exposures in the three gratings were, respectively, 12.3, 8.0, and 4.2 hours, to achieve a uniform $\sim5\sigma$ emission-line sensitivity of $10^{-18}\mbox{ erg}\mbox{ s}^{-1}\mbox{ cm}^{-2}$. In practice, the on-sky performance was roughly a factor of two better. We used a 3-point nod pattern for each observation, and 3 microshutters were allocated for each MSA ``slit." Two dimensional (2D) NIRSpec data were reduced and one-dimensional (1D) spectra were then optimally extracted, corrected for slit losses, and flux calibrated, both in a relative sense from grating to grating, and on an absolute scale, as described in \citet{shapley2024}. 

Since the primary goal of the AURORA program is to detect faint auroral lines from ionized oxygen, sulfur, and nitrogen for $z\sim 1.5-4.5$ galaxies and therefore infer direct metallicities \citep[e.g.,][]{sanders2024a,curti2023},  we required long NIRSpec integration times, and continuous wavelength coverage from 1--5$\mu$m for robust dust corrections. As these observational parameters are also conducive to obtaining useful data for faint $z>6$ galaxies, such higher-redshift targets also comprised a key component of the AURORA mask-design strategy. In total, 97 galaxies were targeted on the two AURORA masks (one in COSMOS and one in GOODSN), 13 of which had a photometric redshift of $z_{\rm phot}\geq 6$. 
GOODSN-100182 was previously identified as a $z>6$ candidate, on the basis of either being a ``z-dropout" \citep[][$z_{\rm phot}=7.41$]{bouwens2015}, or having a photometric redshift  at $z\sim 7$ \citep[][$z_{\rm phot}=7.43$ and 6.77, respectively]{finkelstein2015,jung2020}. The prior overestimated photometric redshifts from \citet{bouwens2015} and \citet{finkelstein2015} can be explained if the far-UV photometric color is interpreted as arising from  an intrinsically bluer SED and the effects of greater Ly$\alpha$ forest blanketing than in the actual spectroscopic solution. Furthermore, it is notable that, despite its remarkable properties revealed by {\it JWST}, GOODSN-100182 was previously detected in the H band at a level of $m_{\rm F160W}\sim 26$~AB, and was therefore not actually missing from the {\it HST} census of $z\sim 7$ galaxies.

\section{Results}
\label{sec:results}

\begin{table}
 \centering
 \caption{Properties of GOODSN-100182
 }\label{tab:emline}
 \begin{tabular}{ l r}
 \hline\hline
%Line  & Flux\\
%\hline
%\hline
\multicolumn{2}{c}{Single-Component Emission-line Fits}\\
\hline\hline
Line  & Flux\tablenotemark{a}\\
\hline
$\mbox{[OII]}\lambda\lambda3726,3729$ & $5.33 \pm 0.16$ \\
$\mbox{[NeIII]}\lambda3869$           & $0.59 \pm 0.11$ \\
H$\delta$                      & $0.83 \pm 0.12$ \\
H$\gamma$                      & $1.89 \pm 0.14$ \\
$\mbox{[FeII]}\lambda4360$    &   $0.53 \pm 0.16$\\
H$\beta$                      & $4.07 \pm 0.13$ \\
$\mbox{[OIII]}\lambda4959$    & $4.55 \pm 0.15$ \\
$\mbox{[OIII]}\lambda5007$    & $13.57 \pm 0.20$ \\
$\mbox{HeI}\lambda5017$    & $0.77 \pm 0.13$ \\
$\mbox{HeI}\lambda5877$    & $0.70 \pm 0.13$ \\
$\mbox{[OI]}\lambda6300$    & $0.87 \pm 0.17$ \\
$\mbox{[NII]}\lambda6548$    & $2.50 \pm 0.18$ \\
H$\alpha$                     & $19.65 \pm 0.30 $ \\
$\mbox{[NII]}\lambda6585$   & $7.13 \pm 0.20$ \\
\hline
\hline
\multicolumn{2}{c}{Double-Component Fit to H$\alpha$}\\
\hline\hline
Line  & Flux\tablenotemark{a}\\
\hline
$\mbox{[NII]}\lambda6548$  (narrow)   & $1.80 \pm 0.53$ \\
H$\alpha$ (narrow)                      & $17.24 \pm 1.54$ \\
H$\alpha$ (broad)                      & $6.23 \pm 0.95$ \\
$\mbox{[NII]}\lambda6585$  (narrow)   & $5.31 \pm 0.78$ \\
\hline\hline
\multicolumn{2}{c}{Galaxy Properties}\\
\hline\hline
Property & Value\\
\hline
R.A. (J2000) & 12:36:42.5184 \\
Decl. (J2000) & $+$62:17:29.4620\\
$z$ & 6.7318\\
$M_{\rm UV}$ & $-20.28\pm 0.06$\\
$\beta$ & $-0.50\pm 0.09$ \\
$\log(M_*/M_{\odot})$ & $9.97^{+0.18}_{-0.24}$\\
$E(B-V)_{\rm gas}$ & $0.40^{+0.10}_{-0.09}$ \\
$\log({\rm SFR(H}\alpha)/ {\rm M}_{\odot}{\rm yr}^{-1})$ & $2.02^{+0.13}_{-0.14}$\\
$r_{\rm eff}$/kpc (multi-component)\tablenotemark{b} & $1.32\pm 0.03$\\
$r_{\rm eff}$/kpc (single)\tablenotemark{c} & $1.63\pm 0.02$\\

\hline
 \end{tabular}
 \tablenotetext{a}{Observed emission-line flux in units of $10^{-18}\mbox{ ergs s}^{-1}\mbox{ cm}^{-2}$. All features here, with the exception of [FeII]$\lambda4360$, are detected at $\geq5\sigma$ significance. [FeII]$\lambda4360$ is detected at $3.3\sigma$ significance, and is listed here to accompany the corresponding discussion in the text.}
 \tablenotetext{b}{Effective radius of disk-component-only in multi-component fit to the light distribution of GOODSN-100182.}
 \tablenotetext{c}{Effective radius of single-component fit to the light distribution of GOODSN-100182.}
\end{table}

\begin{figure}
\centering
\includegraphics[width=1.0\linewidth]{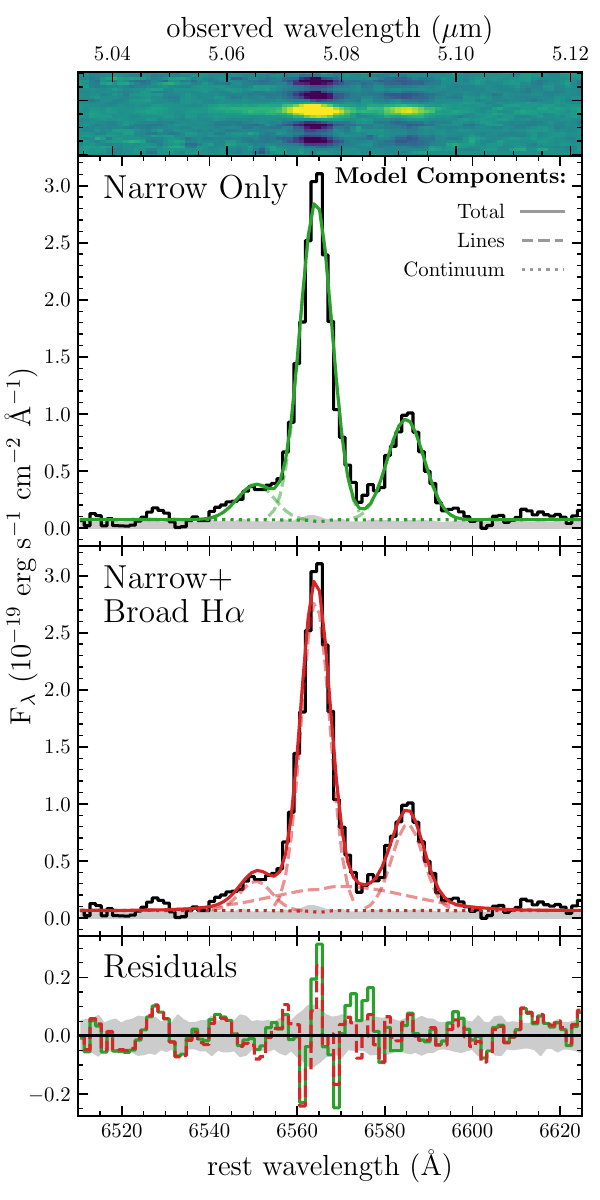}
\caption{H$\alpha$ profile fits. The top panel shows the 2D spectrum of GOODSN-100182 zoomed in on the wavelengths of H$\alpha$ and [NII]$\lambda\lambda6548,6583$ emission. The next panel down features the 1D extracted spectrum of H$\alpha$ and [NII] emission (black curve) along with a single-velocity-component fit to the spectrum (``Narrow Only"; green curve). The spectrum uncertainty is shown as a grey shaded region. The next panel down shows the same 1D extracted spectrum and uncertainty along with a narrow+broad-velocity component fit to H$\alpha$ emission (``Narrow+Broad H$\alpha$"; red curve). The bottom panel plots the residuals from both the ``Narrow Only" and ``Narrow+Broad H$\alpha$" model fits, along with the spectrum uncertainty.
}
\label{fig:ha-fit}
\end{figure}

\subsection{The Spectrum of GOODSN-100182}
\label{sec:results-spectrum}
As shown in Figure~\ref{fig:spec-100182}, the NIRSpec spectrum of GOODSN-100182 contains detections of many rest-frame optical emission lines, including 15 features detected between the wavelengths of [OII]$\lambda\lambda3726,3729$ and [NII]$\lambda6583$. Given the redshift of 100182, most of these lines fall within the G395M grating spectrum. The shortest-wavelength rest-optical features, [OII]$\lambda\lambda3726,3729$ and [NeIII]$\lambda 3869$, fall within the coverage of G235M. 
As described in \citet{shapley2024}, emission lines
were fit using Gaussian profiles, including single peaks for widely-separated lines (e.g., H$\beta$), and multiple simultaneously-fit peaks for closely-spaced or blended lines (e.g., [OII]$\lambda\lambda3726,3729$, and [NII]$\lambda\lambda6548,6583$, and H$\alpha$). 

Furthermore, as Figure~\ref{fig:ha-fit} reveals, the profile of H$\alpha$ emission is not adequately fit by a single velocity component. A better fit is obtained by decomposing the profile into a narrow systemic component with a velocity FWHM of $v_{\rm FWHM}=312 \mbox{ km s}^{-1}$ and an additional, fainter, broad H$\alpha$ component, with both a significantly larger FWHM ($v_{\rm FWHM,broad}=1625 \mbox{ km s}^{-1}$) and an offset in centroid velocity of $+270 \mbox{ km s}^{-1}$.

In Table~\ref{tab:emline}, we report the fluxes of detected emission lines using single-component Gaussians (which provide an adequate description of all lines except H$\alpha$). In addition, we report line fluxes resulting from decomposing H$\alpha$ into  narrow and broad components (broad H$\alpha$, and narrow H$\alpha$, [NII]$\lambda6548$, and [NII]$\lambda 6583$). In the emission-line ratio analysis and plots that follow, results are plotted for 100182 using both single-component fits for H$\alpha$ and [NII], as well as the narrow component of the more complex fit. The observed centroid wavelengths of the emission lines in the spectrum of GOODSN-100182  indicate a redshift of $z_{\rm spec}=6.7318$.  

\subsection{The Physical Properties of GOODSN-100182}
\label{sec:results-props}
The combination of {\it JWST}/NIRCam imaging and NIRSpec spectroscopy provides a window into an exceptional $z\sim 7$ galaxy, including its spectral energy distribution (SED) and inferred stellar population parameters, its structural properties and size, its dust content, current SFR, and chemical abundance. In Table~\ref{tab:emline}, we summarize some of the basic properties of GOODSN-100182 that are described in the following section.

\begin{figure}
\centering
\includegraphics[width=1.0\linewidth]{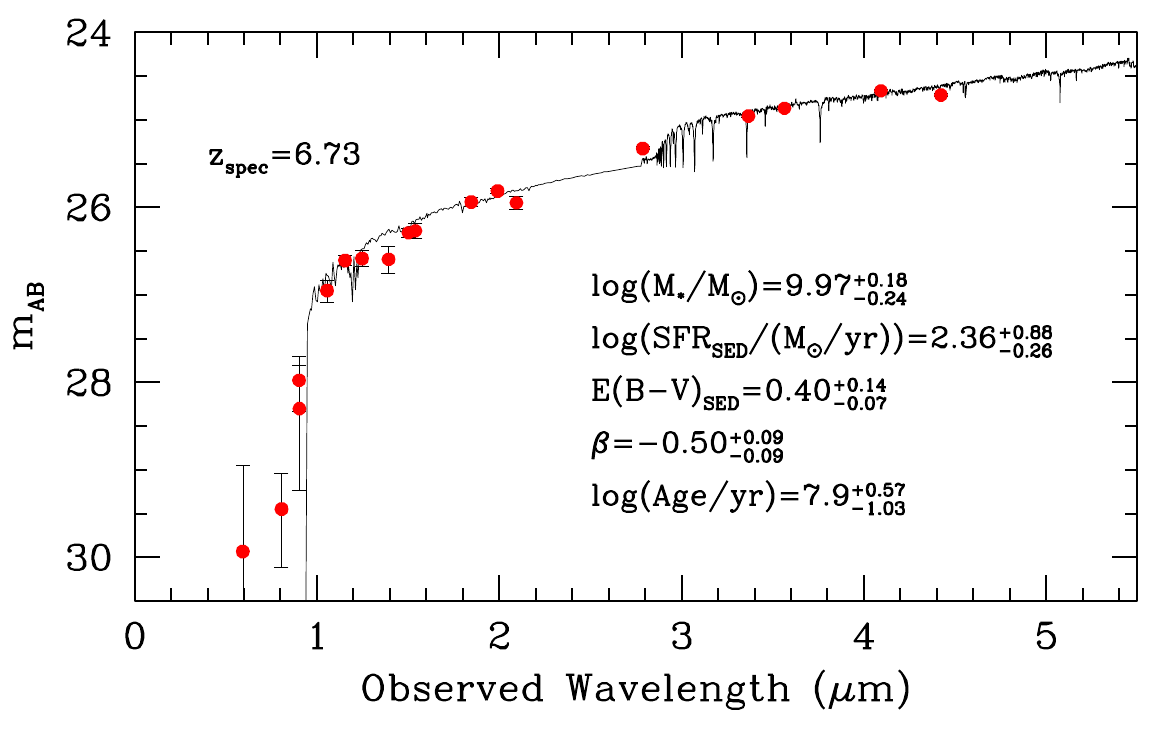}
\caption{Stellar population of GOODSN-100182.
The observed and best-fit model SEDs for GOODSN-100182. Photometric detections are indicated with red symbols, and have been corrected for the contributions of nebular emission lines and continuum. The best-fit stellar population model of GOODSN-100182 is shown as the solid black curve. We also list the parameters of this model from \citet{conroy2009}, which assumes solar metallicity and a Calzetti attenuation law.}
\label{fig:sed-100182}
\end{figure}

\begin{figure*}
\centering
\includegraphics[width=1.0\linewidth]{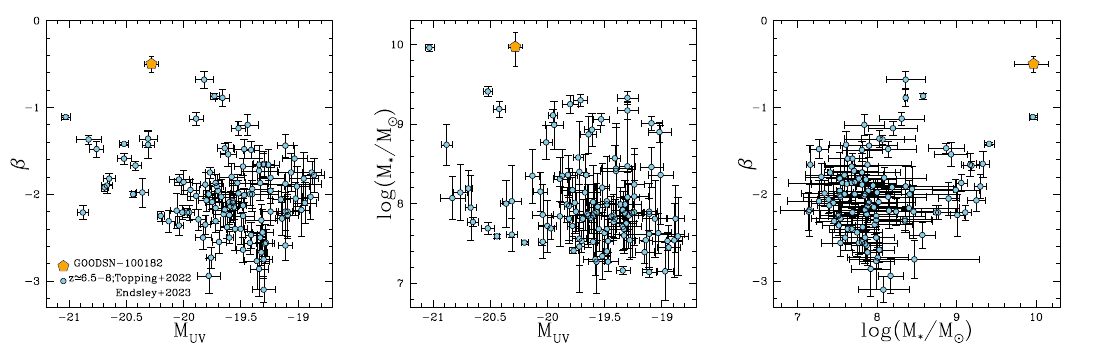}
\caption{Photometric properties of GOODSN-100182. In each panel, GOODSN-100182 is indicated using an orange pentagon, while the comparison sample of $z\sim 6.5-8$ star-forming galaxies from \citet{topping2022} and \citet{endsley2023} is shown using light blue circles. {\bf Left:} UV slope ($\beta$) vs. $M_{\rm UV}$. GOODSN-100182 is redder than all galaxies in the comparison sample, including at fixed $M_{\rm UV}$. {\bf Center:} Stellar mass ($\log(M_*/M_{\odot})$) vs. $M_{\rm UV}$. GOODSN-100182 is more massive than all galaxies in the comparison sample, including at fixed $M_{\rm UV}$. {\bf Right:} 
 UV slope ($\beta$) vs. $\log(M_*/M_{\odot})$. While overlapping with the comparison sample in $M_{\rm UV}$, GOODSN-100182 clearly lies at an extreme of the $z\sim 7$ star-forming population in terms of UV color and stellar mass.}
\label{fig:betamuvmstar}
\end{figure*}

\begin{figure*}
\centering
\includegraphics[width=0.97\linewidth]{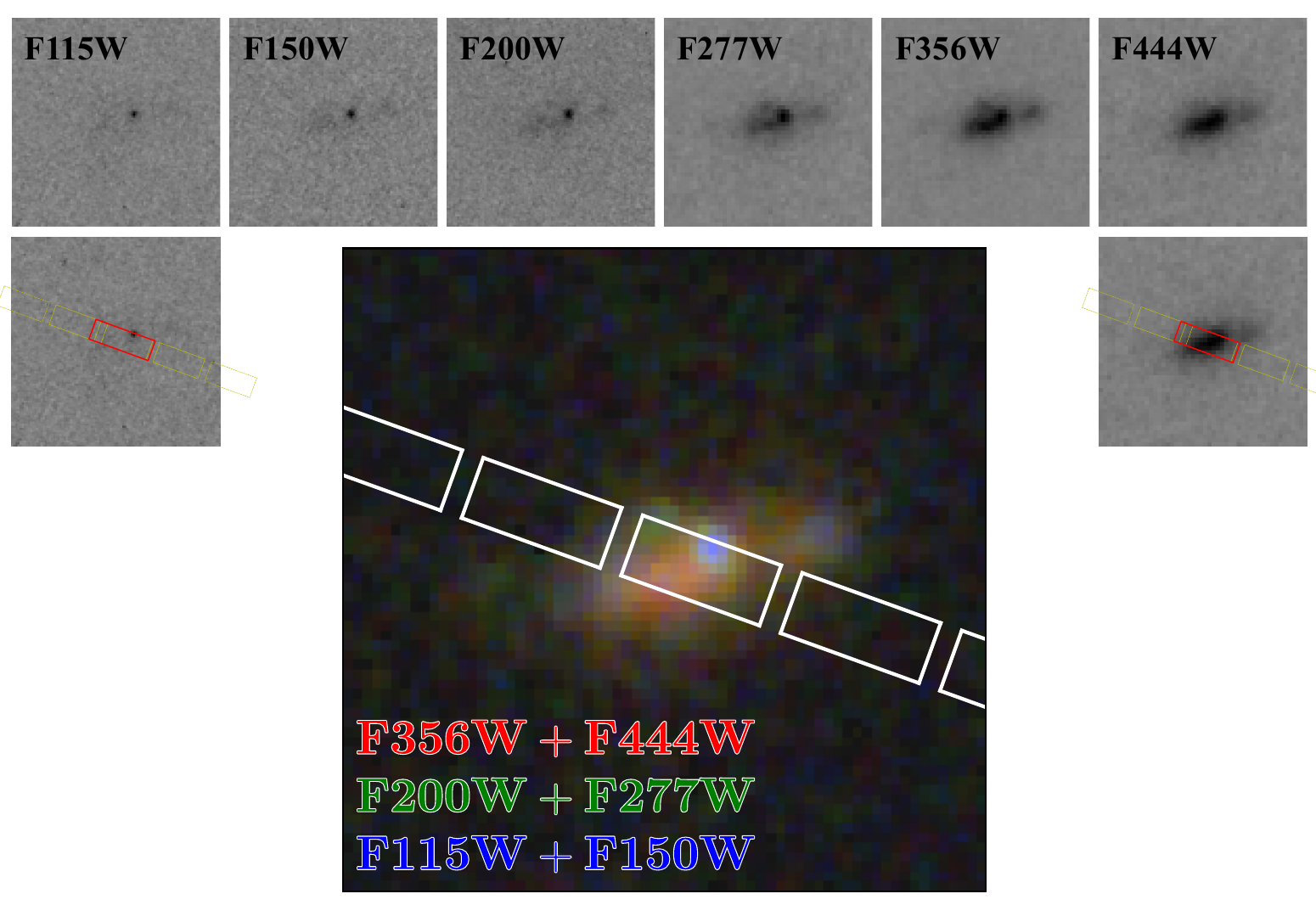}
\caption{NIRCam Images of GOODSN-100182. Postage stamps are shown for NIRCam F115W, F150W, F200W, F277W, F356W, and F444W. Each postage stamp is 2"$\times$2" and shown with North up and East to the left. The F115W and F444W postage stamps are duplicated with the NIRSpec slit overlaid in yellow, in which the extraction window is highlighted in red. In addition,  we show a 2"$\times$2" colormap
of GOODSN-100182, where F115W+F150W, F200W+F277W, and F356W+F444W comprise, respectively, the blue, green, and red channels. The blue (in a relative sense) point source is only partially covered by the NIRSpec slit.}
\label{fig:stamps-100182}
\end{figure*}

\subsubsection{Spectral Energy Distribution}
\label{sec:results-props-sed}
We can model the SED of GOODSN-100182 in order to infer its stellar population properties. As this galaxy lies in the GOODSN field, it is covered by an extensive suite of multi-wavelength imaging. We used the publicly-available GOODSN photometric catalog in the DAWN {\it JWST} Archive\footnote{https://dawn-cph.github.io/dja/} \citep{heintz2024}, which includes {\it HST}/ACS and WFC3 F435W, F606W, F775W, F814W, F850LP, F105W, F125W, F140W, F160W data from various programs and assembled in \citet{skelton2014}, and NIRCam imaging obtained through the JADES, FRESCO, and JEMS programs spanning F090W, F115W, F150W, F182M, F200W, F210M, F277W, F335M,  F356W, F410M, and F444W
\citep{eisenstein2023,oesch2023,williams2023}. 
The photometry was corrected for the contributions from nebular emission lines using the method described in \citet{sanders2021}, and  additionally corrected for contributions from nebular continuum emission based on predictions from grids of Cloudy photoionization models \citep{ferland2017}, tied to the measured H$\beta$ line flux \citep{sanders2024b}.

For our fiducial model, we used the program FAST \citep{kriek2009}, with the assumption of
the flexible stellar population synthesis (FSPS) models of \citet{conroy2009}, solar metallicity,  a \citet{chabrier2003} IMF, and the \citet{calzetti2000} dust attenuation for the reddening of the stellar continuum. The star-formation history was parametrized using the ``delayed-$\tau$" form, i.e., ${\rm SFR}(t)\propto t \times exp(-t/\tau)$. Here, $t$ is time since the onset of star formation and $\tau$ is
the characteristic star-formation timescale.
As shown in Figure~\ref{fig:sed-100182}, the best-fit parameters are $\log({\rm M}_*/{\rm M}_{\odot})=9.97^{+0.18}_{-0.24}$, $\log({\rm SFR}/ {\rm M}_{\odot}{\rm yr}^{-1})=2.36^{+0.88}_{-0.26}$, 
%$A_V=1.60{+0.56}_{-0.29}$, 
$E(B-V)_{\rm SED}=0.40^{+0.14}_{-0.07}$,
$\log(t/{\rm yr})=7.90^{+0.57}_{-1.03}$, $\log(\tau/{\rm yr})=8.2^{+1.8}_{-0.2}$. For completeness, we also modeled the SED of GOODSN-100182 using \textsc{prospector} \citep{johnson2021}, assuming a non-parametric star-formation history (SFH). For this model setup, we assumed a SFH comprising eight independent time bins constrained by a continuity prior \citep[see, e.g.,][]{tacchella2022}. Notably, the \textsc{prospector} fit also yields a high stellar mass, with $\log({\rm M}_*/{\rm M}_{\odot})=10.15 \pm 0.10$. In fact, the large stellar mass of GOODSN-100182 was already suggested prior to the launch of {\it JWST}, based on its {\it HST}+{\it Spitzer}/IRAC SED, from which \citet{jung2020} inferred $\log({\rm M}_*/{\rm M}_{\odot})=10.20^{+0.05}_{-0.08}$ \citep[][Jung et al., private communication]{jung2020}.

A common empirical diagnostic for characterizing the SED shape of star-forming galaxies is $\beta$, the UV slope. We estimate  $\beta$ for GOODSN-100182 by fitting a power law of the form $f_{\lambda} \propto \lambda^{\beta}$ to the photometric bands spanning the rest-wavelength range $1300-2600$\AA\ (i.e., F105W, F115W, F125W, F140W, F150W, F160W, and F200W). Based on the SED of 100182, we find $\beta=-0.50 \pm 0.09$, significantly redder than the median slope of $\beta=-2.0$ for  $z\sim 7$ star-forming galaxies in \citet{topping2022}. In fact, there are no galaxies with such red slopes in the $z\sim 7$ sample of \cite{topping2022}. 

Taken together, the basic photometric properties of GOODSN-100182 are atypical for $z\sim 7$ star-forming galaxies. We use the sample presented in \citet{topping2022} and \citet{endsley2023} for context. As shown in Figure~\ref{fig:betamuvmstar} (left), in the space of $\beta$ vs. $M_{{\rm UV}}$, where ``UV" is defined as rest-frame 1500~\AA\ and is $-20.28 \pm 0.06$ for GOODSN-100182, this galaxy stands out as being redder than all other galaxies at fixed $M_{{\rm UV}}$. 
Correspondingly, GOODSN-100182 has a significantly higher stellar mass than the average $z\sim 7$ star-forming galaxy, at fixed $M_{{\rm UV}}$ (Figure~\ref{fig:betamuvmstar}, center; see also \citealt{stefanon2021}). 
Finally, it is clear that GOODSN-100182 occupies an extreme of the distribution of $z\sim 7$ star-forming galaxies, as its stellar mass is higher and its UV slope is redder than any galaxy in the comparison sample (Figure~\ref{fig:betamuvmstar}).

\begin{figure}
\centering
\includegraphics[width=1.0\linewidth]{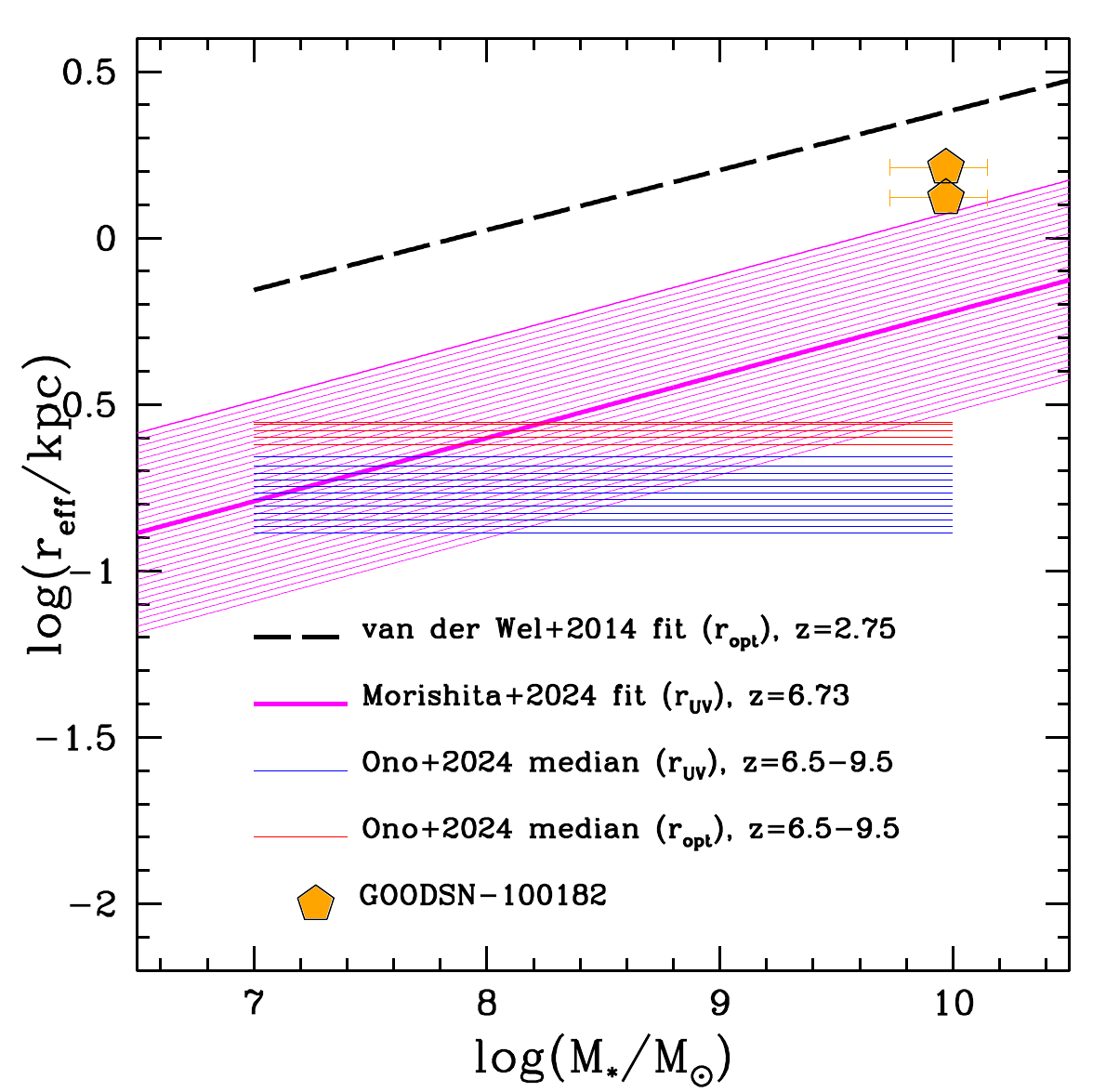}
\caption{Size ($r_{\rm eff}$) vs. stellar mass. GOODSN-100182 is plotted as an orange star, with two $r_{\rm eff}$ values ($r_{\rm eff}=1.32$~kpc in the multi-component S\'ersic fit, and 1.63~kpc in the single-component S\'ersic fit). The size-mass relation of \citet{morishita2024a} tuned to $z=6.73$ is shown as a solid magenta curve, and the intrinsic $1\sigma$ scatter as a shaded magenta region. We also include the range of median rest-UV (blue shaded region) and rest-optical (red shaded region) sizes from \citet{ono2024} for $z=6.5-9.5$ galaxies with comparable 
$M_{\rm UV}$ to GOODSN-100182. The horizontal extent of the shaded regions corresponds to the range of stellar masses probed in that sample. GOODSN-100182 lies closer to the $z=2.75$ size-mass relation from \citet{vanderwel2014} than to the parameters describing typical $z\sim 7$ star-forming galaxies.}
\label{fig:sizevmass}
\end{figure}

\subsubsection{Structural Properties}
\label{sec:results-props-structure}
{\it JWST}/NIRcam provides the opportunity to investigate the structural properties of GOODSN-100182 at rest-UV through rest-optical wavelengths in unprecedented detail.
Figure~\ref{fig:stamps-100182} shows 2"$\times$2" images of the galaxy in the F115W, F150W, F200W, F277W, F356W, and F444W filters, spanning from rest-frame $1500 -5700$\AA, along with an RGB colormap (B=F115W$+$F150W; G=F200W$+$F277W; R=F356W$+$F444W). We also show overlays of the NIRSpec slit on the F115W and F444W images, and the colormap. 

Qualitatively, it is clear that GOODSN-100182 is both spatially extended -- as most strikingly revealed in the rest-frame optical by the new {\it JWST}/NIRCam imaging -- and has a point-like source towards the center of the light distribution that is more prominent at shorter wavelengths. We followed the methodology in \citet{pahl2022} to quantitatively determine the structural properties of GOODSN-100182. 
Specifically, we used the software \textsc{galfit} \citep{peng2002,peng2010} to fit one or more S\'ersic profiles and an unresolved ``PSF" component. 

Most ($\geq 80$\%) of the light in the rest-optical (F356W and F444W) is contained in a S\'ersic component with $n\sim 1$, with minor ($5-10$\%)  additional contributions from an unresolved component towards the center of the galaxy, and an additional, resolved surface-brightness concentration towards the western edge of the galaxy ($10-15$\%). At shorter wavelengths (e.g., F115W) the western component is not present, and the unresolved component is more prominent (contributing at the $\sim 15$\% level to the total flux). The colormap of GOODSN-100182 reflects the increased prominence of the unresolved component at shorter wavelengths, in that this surface-brightness peak appears relatively blue compared to its surroundings. 

In multi-component fits to the rest-optical light distribution, the dominant disk component is described by an effective radius along the semi-major axis of $r_{\rm eff}=1.32\pm 0.03$~kpc (the value reported is the average for F356W and F444W). If, on the other hand, we use a single-component S\'ersic fit to the GOODSN-100182 light distribution, we find $r_{\rm eff}=1.63\pm 0.02$~kpc (and still $n\sim 1$).
As shown in Figure~\ref{fig:sizevmass}, GOODSN-100182 is a factor of $\sim 5$ larger in effective radius than the median for star-forming galaxies with comparable UV luminosities at $z\sim 7$ \citep{ono2024}. It is also a factor of $\sim 2$ larger than the median of a sample of $10^{10} M_{\odot}$ $z\sim 7$ galaxies drawn from the CEERS survey \citep{chworowsky2024}.

\begin{figure}
\centering
\includegraphics[width=1.0\linewidth]{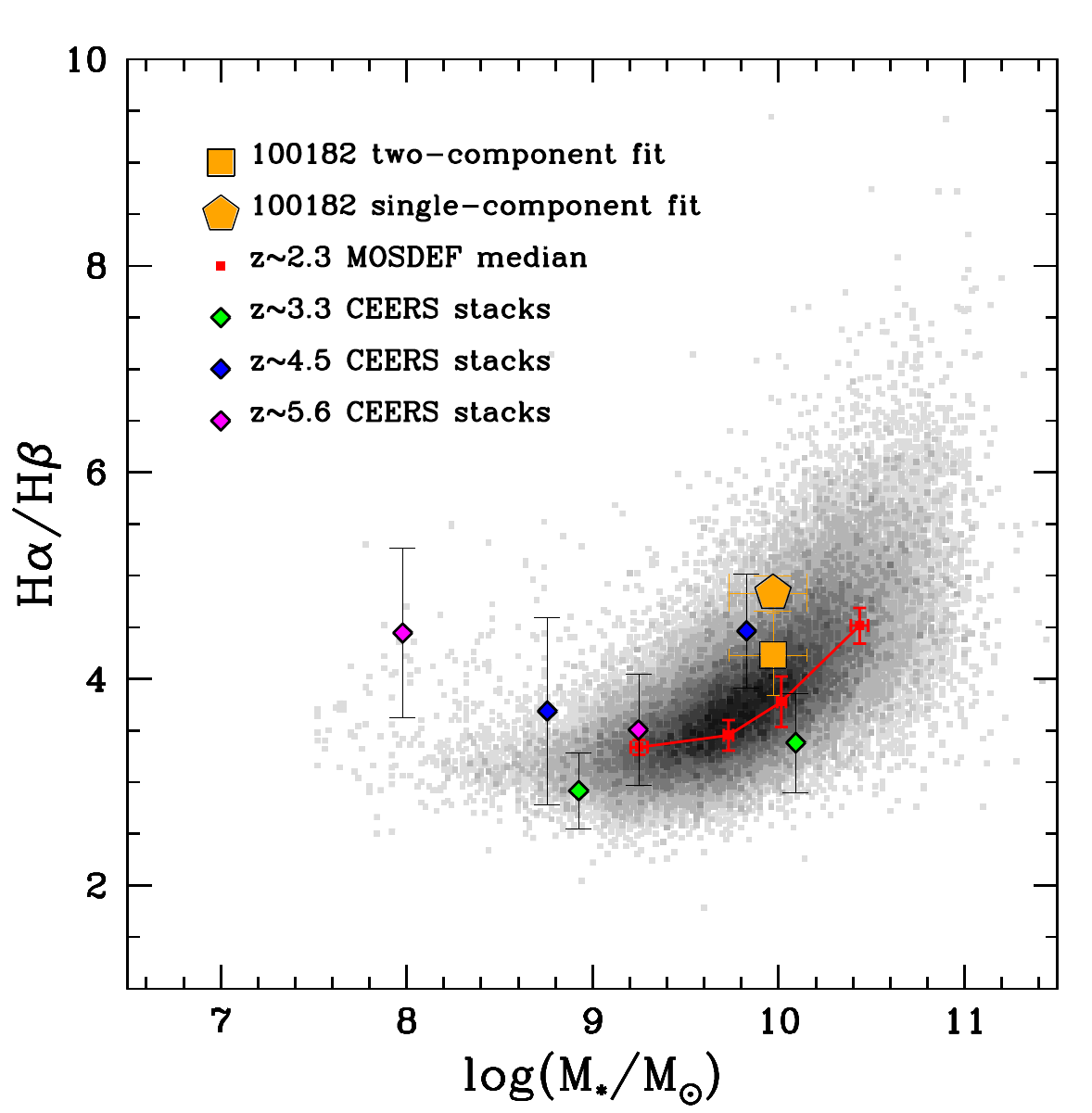}
\caption{Attenuation vs. $M_*$ based on the Balmer line ratio, H$\alpha$/H$\beta$. Here and in all subsequent emission-line plots, GOODSN-100182 is plotted as an orange symbol (the square shows the results for the two-component fit to H$\alpha$, whereas the pentagon represents the case of the single-component H$\alpha$ fit). The background greyscale  histogram  here and in all subsequent emission-line plots corresponds to the distribution of local SDSS galaxies. The running median H$\alpha$/H$\beta$ ratio for $z\sim 2.3$ star-forming galaxies in the MOSDEF survey is shown in red \citep{shapley2022}. In addition, H$\alpha$/H$\beta$ ratios are measured from composite spectra drawn from the CEERS survey in two bins of stellar mass for each of three redshift ranges, where $\langle z\rangle=3.3$ (green diamonds), $\langle z\rangle=4.5$ (blue diamonds); and $\langle z\rangle=5.6$ (magenta diamonds) \citep{shapley2023b}.
}
\label{fig:hahblm}
\end{figure}

\begin{figure*}
\centering
\includegraphics[width=1.0\linewidth]{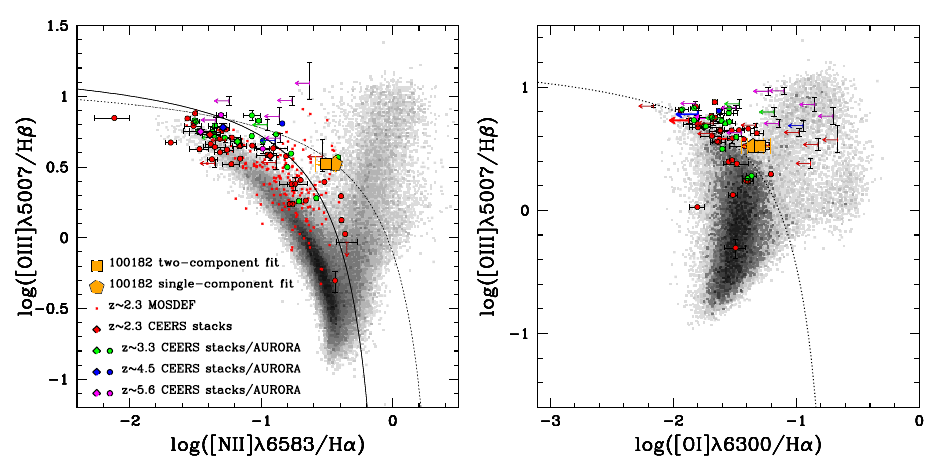}
\caption{``BPT" Emission-line Diagnostic Diagrams. GOODSN-100182 and emission-line ratios from CEERS composite spectra are plotted using the same symbols as in Figure~\ref{fig:hahblm}.
 In each panel, individual galaxies from the AURORA sample are plotted with colored symbols. Red, green, blue, and magenta are used, respectively, for the $1.4 \leq z < 2.7$, $2.7\leq z < 4.0$, $4.0 \leq z < 5.0$, and $z\geq 5.0$ samples. Galaxies with $\geq3\sigma$ detections of all 4 BPT emission lines are plotted with circles, while 3$\sigma$ limits are shown as arrows. Individual $z\sim 2.3$ galaxies from the MOSDEF survey  (where available) are plotted with small red points. {\bf Left:} [NII] BPT diagram. The  ``maximum starburst" line from \citet{kewley2001} is shown as a dotted curve,  while the empirical AGN/star-formation threshold from \citet{kauffmann2003} drawn as a black solid curve {\bf Right:} [OI] BPT diagram. The black dotted
curve is the ``maximum starburst" line from \citet{kewley2001}.  }
\label{fig:bpt}
\end{figure*}

\subsubsection{Line Ratios: Dust, Star Formation, Excitation, and Metallicity}
\label{sec:results-props-lineratio}
The rich spectrum of GOODSN-100182 provides a detailed view of its ionized star-forming ISM, including its dust content, instantaneous rate of star formation, and excitation and chemical enrichment.  

\paragraph{Dust Attenuation}
\label{sec:results-props-lineratio-dust}
We begin with dust and star formation. The nebular dust attenuation is estimated from the H$\alpha$/H$\beta$ flux ratio using the narrow component of H$\alpha$. Assuming an intrinsic ratio of 2.86 for H$\alpha$/H$\beta$ and the \citet{cardelli1989} attenuation curve, we find $E(B-V)_{\rm gas}=0.40^{+0.10}_{-0.09}$. The $E(B-V)_{\rm gas}$ inferred from the H$\alpha$/H$\gamma$ and H$\alpha$/H$\delta$ ratios is consistent with that inferred from H$\alpha$/H$\beta$. All Balmer ratios suggest significant dust attenuation. GOODSN-100182 can also be considered in the context of the relationship between attenuation and stellar mass, which does not evolve strongly between $z\sim0$ and $z\sim 6$. In this context, GOODSN-100182 is characterized by a higher than average H$\alpha$/H$\beta$ ratio, given its stellar mass (Figure~\ref{fig:hahblm}), where the average is similar for both $z\sim 0$ galaxies from the Sloan Digital Sky Survey \citep[SDSS;][]{abazajian2009} and samples at $z\geq 2$ \citep{shapley2022,shapley2023b,sandles2024}.

\begin{figure*}
\centering
\includegraphics[width=1.0\linewidth]{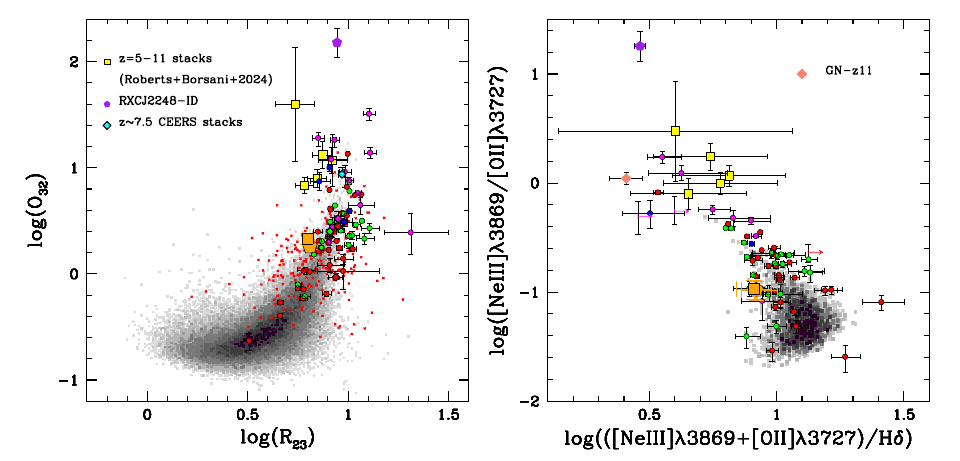}
\caption{Ionization-excitation diagrams.
{\bf Left:} $O_{32}$ vs. $R_{23}$ diagram, corrected for dust. Symbols for GOODSN-100182, AURORA and MOSDEF galaxies, and CEERS composites, are as in Figure~\ref{fig:bpt}. We also show the CEERS $\langle z\rangle \sim 7.5$ composite (cyan diamond), RXCJ2248-ID ($z=6.1$, purple pentagon) from \citet{topping2024}, and the stacked spectra at $\langle z \rangle=5.5-9.5$ (yellow squares) from \citet{robertsborsani2024}. The $z\sim 0$ SDSS histogram includes only galaxies identified as star-forming using the criteria of \citet{kauffmann2003}, given that AGNs do not clearly segregate in this diagram. {\bf Right:} Dust-corrected [NeIII]$\lambda 3869$/[OII]$\lambda3727$ vs. ([NeIII]$\lambda3869$+[OII]$\lambda3727$)/H$\delta$. Symbols as in the left-hand panel, with the addition of GN-z11 ($z=10.6$, pink diamond) from \citet{bunker2023}.
}
\label{fig:ion-exc}
\end{figure*}

\paragraph{Star formation}
\label{sec:results-props-lineratio-sfr}
Based on the value of $E(B-V)_{\rm gas}$ and the observed H$\alpha$ flux, and assuming the H$\alpha$ to SFR calibration from \citet{hao2011} normalized to a \citet{chabrier2003} IMF, we estimate the instantaneous star-formation rate to be 
$\log({\rm SFR(H}\alpha)/ {\rm M}_{\odot}{\rm yr}^{-1})=2.02^{+0.13}_{-0.14}$. Notably, this value is consistent with the SFR inferred from SED fitting. Compared to recent estimates of the star-forming main sequence at $z\sim 6-7$ \citep{cole2023,clarke2024}, GOODSN-100182 has a significantly higher than average (factor of $\sim 3$) specific SFR. However, there are very few galaxies sampling the main sequence at $M_*\geq 10^{10} M_{\odot}$, so most of the constraining power for the form of the main sequence at this redshift comes from lower-mass systems than GOODSN-100182.

\paragraph{Excitation and Metallicity}
\label{sec:results-props-lineratio-excmet}
It is also possible to investigate the excitation and metallicity of the ionized ISM in GOODSN-100182 using several diagnostic line emission-line ratios. To place GOODSN-100182 in context, we use  both local SDSS galaxies and low-mass systems from the Local Volume Legacy (LVL) Survey \citep{berg2012}, and also samples at $1.4\leq z \leq 3.8$ from the MOSDEF survey \citep{kriek2015}, at $2.0 \leq z \leq 9.3$ from the CEERS survey \citep{finkelstein2023,sanders2023b,shapley2023a},  at $1.4\leq z \leq 7.5$ from the parent AURORA sample \citep{shapley2024}, the $z=5-11$ composite spectra of \citet{robertsborsani2024}, and notable individual $z>6$ galaxies. None of our key results is sensitive to whether we assume one-component or two-component fits to the H$\alpha$ of GOODSN-100182. However, for completeness, in all plots we indicate the location of GOODSN-100182 assuming both types of profile fit.

Starting with the ``BPT" diagrams of \citet{baldwin1981} and \citet{veilleux1987}, we find that, in the space of [OIII]$\lambda5007$/H$\beta$ vs. either [NII]$\lambda6583$/H$\alpha$ or [OI]$\lambda6300$/H$\alpha$, GOODSN-100182 overlaps with the properties of star-forming $z\sim 2-3$ galaxies (Figure~\ref{fig:bpt}). More specifically, in the [NII]$\lambda6583$/H$\alpha$ BPT diagram, the line ratios of GOODSN-100182 are most similar to those of the most massive, metal-rich systems in the MOSDEF sample \citep{shapley2019}. Furthermore, like $z\sim 2$ MOSDEF galaxies, GOODSN-100182 is offset towards higher [NII]$\lambda6583$/H$\alpha$ and/or [OIII]$\lambda5007$/H$\beta$ relative to the locus of $z\sim 0$ star-forming galaxies. As discussed in many previous works \citep[e.g.,][]{steidel2016,shapley2019,runco2021}, this offset can arise due to chemically-young, $\alpha$-enhanced stellar populations, with harder ionizing spectra at fixed nebular oxygen abundance. The inferred short timescale for star formation in GOODSN-100182 ($\leq 100$~Myr) is consistent with the presence of $\alpha$-enhanced massive stars.

Another key set of diagnostic diagrams probe both ionization parameter and metallicity. Traditionally, these properties have been studied with the  O$_{32}$ ([OIII]$\lambda5007$/[OII]$\lambda3727$) vs. R$_{23}$ (([OII]$\lambda3727$+[OIII]$\lambda\lambda4959,5007$))/H$\beta$) diagram, in which O$_{32}$ traces ionization parameter, and R$_{23}$ is sensitive to metallicity. The O$_{32}$ vs. R$_{23}$ diagram has now been measured for galaxies at $z\geq 9$ \citep{sanders2023b,cameron2023,robertsborsani2024}, showing a general progression towards higher  O$_{32}$ values at earlier times (Figure~\ref{fig:ion-exc}, left). At $z>10$, however, [OIII]$\lambda\lambda4959,5007$ is no longer accessible to {\it JWST}/NIRSpec. In order to investigate the excitation conditions in GN-z11 ($z=10.6$), \citet{bunker2023} introduced an alternative ionization-metallicity diagram based on emission lines at shorter rest-frame wavelengths ([NeIII]$\lambda3870$/[OII]$\lambda3727$ vs. ([NeIII]$\lambda3870$+[OII]3727)/H$\delta$). Here, [NeIII]$\lambda3870$/[OII]$\lambda3727$ serves the same role as O$_{32}$ as a probe of ionization parameter, while ([NeIII]$\lambda3870$+[OII]3727)/H$\delta$ serves as an analog of R$_{23}$ in tracing metallicity. With decreasing metallicity, [NeIII]$\lambda3870$/[OII]$\lambda3727$ tends to increase while ([NeIII]$\lambda3870$+[OII]3727)/H$\delta$ decreases (Figure~\ref{fig:ion-exc}, right). GN-z11 and the lensed $z=6.1$ star-forming galaxy RXCJ-2248ID \citep{topping2024} lie towards one extreme of the distribution, with evidence for high ionization parameter and low metallicity. Remarkably, in both of these diagrams, GOODSN-100182 at $z=6.73$ falls in the region occupied by star-forming galaxies at $z\sim 2-3$, with significantly lower O$_{32}$ and [NeIII]$\lambda3870$/[OII]$\lambda3727$ than measured from the stacked $\langle z\rangle =6.33$ and $\langle z\rangle =7.47$ spectra from \citet{robertsborsani2024}, GN-z11, RXCJ2248-ID, and the majority of $z>5$ galaxies in the AURORA sample.

\begin{figure*}
\centering
\includegraphics[width=1.0\linewidth]{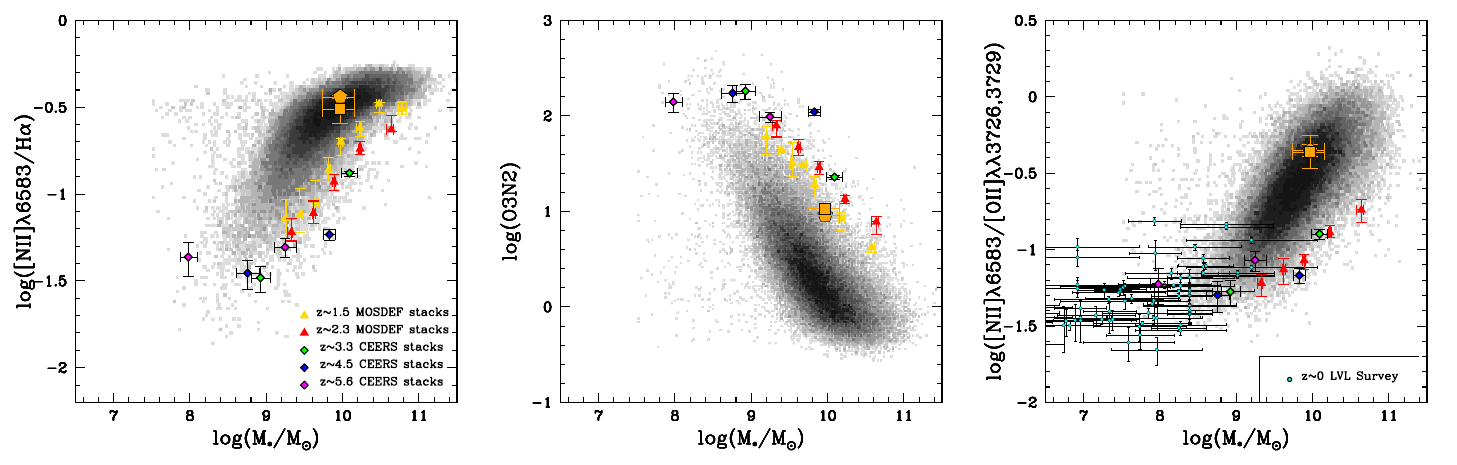}
\caption{Empirical mass-metallicity diagrams. These panels show the emission-line ratios [NII]$\lambda6583$/H$\alpha$ (left), O3N2 (center), and [NII]$\lambda6583$/[OII]$\lambda3727$ (right) versus ${\rm M}_*$. Symbols for GOODSN-100182 are as in previous figures. Composite spectra from the CEERS survey are again shown in two bins of stellar mass for each of three redshift ranges, where $\langle z\rangle=3.3$ (green diamonds), $\langle z\rangle=4.5$ (blue diamonds); and $\langle z\rangle=5.6$ (magenta diamonds). Measurements from stacked spectra of galaxies drawn from the MOSDEF survey at $z\sim 1.5$ \citep{topping2021} and $z\sim 2.3$ \citep{sanders2021} are plotted, respectively, using large gold and red triangles.
In the right panel ([NII]$\lambda6583$/[OII]$\lambda3727$), we plot low-mass galaxies from the Local Volume Legacy (LVL) Survey \citep{berg2012} using small cyan symbols. This sample populates the regime of primary nitrogen enrichment in which there is no dependence of N/O on metallicity.
}
\label{fig:emp-met}
\end{figure*}

\begin{figure}
\centering
\includegraphics[width=1.0\linewidth]{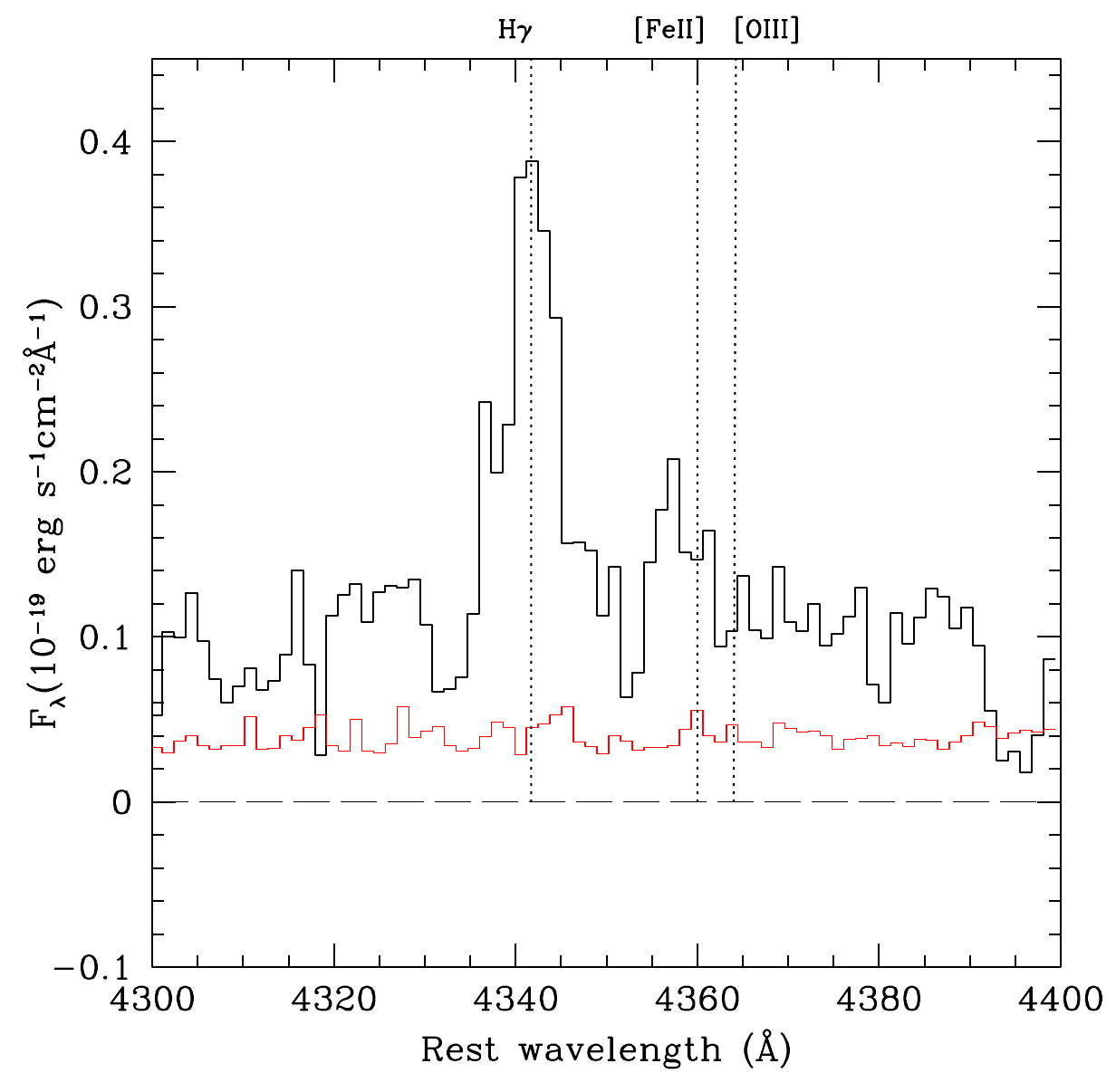}
\caption{Spectrum of H$\gamma$ and [FeII]$\lambda4360$ emission. This figures shows the spectrum of GOODSN-100182 in the vicinity of H$\gamma$ emission. The black and red curves are, respectively,  the science and error spectra. There is additional emission to the red of the H$\gamma$ emission feature, which lines up with the wavelength of [FeII]$\lambda4360$, {\it not} [OIII]$\lambda4363$. As discussed by \citet{curti2017},  the strength of [FeII]$\lambda4360$ emission increases as a strong function of metallicity. Its detection provides independent evidence of the high metallicity of GOODSN-100182.
}
\label{fig:spec-100182-hgamma}
\end{figure}

The rest-optical emission-line spectra of star-forming galaxies also vary systematically as a function of stellar mass, reflecting the underlying dependence of gas-phase metallicity on galaxy stellar mass. Commonly analyzed metallicity-sensitive emission-line ratios include [NII]$\lambda6583$/H$\alpha$, ([OIII]$\lambda5007$/H$\beta$)/[NII]$\lambda6583$/H$\alpha$ (O3N2), or [NII]$\lambda6583$/[OII]$\lambda3727$. Figure~\ref{fig:emp-met} places GOODSN-100182 in the context of these empirical mass-metallicity scaling relations for galaxies spanning a redshift range from $z\sim 0$ to $z\sim 6.5$. As evident in Figure~\ref{fig:spec-100182}, nitrogen emission in GOODSN-100182 is significantly stronger than in typical star-forming $z>5$ galaxies presented in the literature thus far \citep{sanders2023b,shapley2023a}. For example, in \citet{sanders2023b}, the composite spectrum of $z\sim 5.6$ star-forming galaxies from the CEERS survey shows $\log({\rm [NII]/H}\alpha)=-1.46^{+0.12}_{-0.09}$, whereas GOODSN-100182 has $\log({\rm [NII]/H}\alpha)=-0.51^{+0.07}_{-0.08}$. Some of this offset can be attributed to the different stellar mass ranges probed by the CEERS $z\sim 5.6$ composite (median $\log(M_*/M_{\odot}=8.57^{+0.04}_{-0.13}$) and GOODSN-100182 ($\log(M_*/M_{\odot}=9.97^{+0.18}_{-0.24}$). However, it is evident from Figure~\ref{fig:emp-met} that GOODSN-100182 has the same [NII]$\lambda6583$/H$\alpha$ ratio as SDSS $z\sim 0$ galaxies of the same stellar mass, whereas star-forming galaxies from the MOSDEF survey at $z\sim 1.5$ and  $z\sim 2.3$ with $\log(M_*/M_{\odot})\sim 10$ have $\log({\rm [NII]/H}\alpha)$ 0.2 dex lower. In the space of O3N2 vs. stellar mass, GOODSN-100182 is intermediate between the MOSDEF $z\sim 1.5$ and SDSS $z\sim 0$ samples.
Finally in the plot of [NII]$\lambda6583$/[OII]$\lambda3727$, GOODSN-100182 again resembles SDSS $z\sim 0$ galaxies of the same stellar mass, with 
$\log({\rm [NII]}\lambda6583/{\rm [OII]}\lambda3727)$ 0.6 dex higher than MOSDEF $z\sim 2.3$ of similar mass.

The emission-line ratios of GOODSN-100182 span a range that is not covered by recent attempts to calibrate high-redshift galaxy chemical abundances using direct metallicities of high-redshift galaxies, themselves \citep{sanders2024a}.
Therefore, we utilize the metallicity calibrations of \citet{bian2018} for local analogs of $z\sim 2$ star-forming galaxies, in order to translate the ratio of strong nebular emission lines (e.g., [NII]$\lambda6583$/H$\alpha$, O3N2, and O$_{32}$) to oxygen abundance. For [NII]$\lambda6583$/H$\alpha$, we find $12+\log({\rm O/H}_{{\rm N2}})=8.57^{+0.03}_{-0.04}$ and O3N2, we find 
$12+\log({\rm O/H}_{{\rm O3N2}})=8.57^{+0.03}_{-0.03}$. Both line ratios suggest an oxygen abundance that is 75\% solar \citep{asplund2009}. The other line ratio showing a simple linear relationship with metallicity is $O_{32}$, for which the local-analog calibration of \citet{bian2018} yields $12+\log({\rm O/H}_{{\rm O32}})=8.35^{+0.03}_{-0.03}$ (45\% solar). Accordingly, GOODSN-100182 appears to be enriched to a significant fraction of solar metallicity, at $z=6.73$.

As further evidence of advanced chemical enrichment in GOODSN-100182, we zoom in on the spectral region near H$\gamma$ (Figure~\ref{fig:spec-100182-hgamma}). Adjacent to H$\gamma$, we detect emission with a rest-frame centroid of 4359.8\AA\, which we identify as [FeII]$\lambda4360$, {\it not} auroral [OIII]$\lambda 4363$ emission. As discussed by \citet{curti2017}, this [FeII] feature increases in strength relative to [OIII]$\lambda 4363$ with increasing metallicity, dominating over the flux of the [OIII]$\lambda 4363$ feature at roughly solar metallicity. 

\begin{figure*}
\centering
\includegraphics[width=1.0\linewidth]{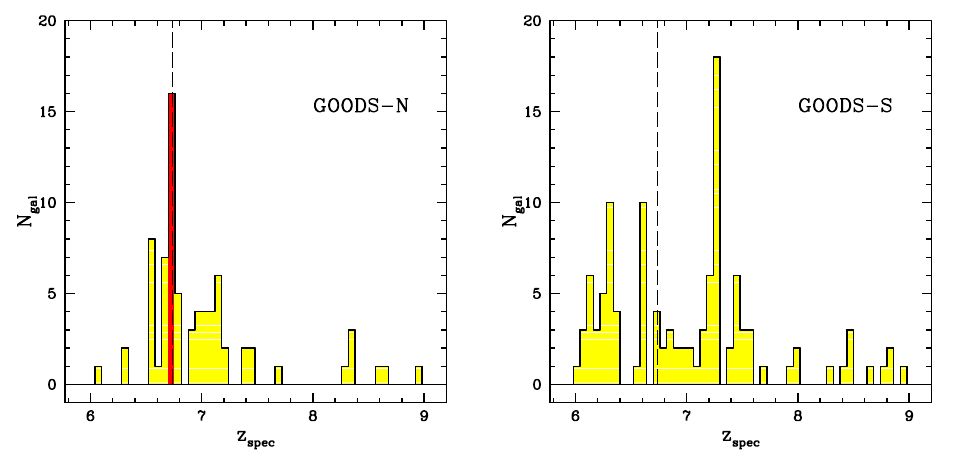}
\caption{Redshift histograms spanning $6\leq z \leq 9$ in the GOODS-N and GOODS-S fields. These spectroscopic redshift histograms are constructed from the JADES Data Release 3 \citep{deugenio2024}, and each bin has $\Delta z=0.06$. {\bf Left:} Redshift histogram in the GOODS-N field. The bin located at $z=6.73\pm 0.03$ is shaded red and marked with a vertical dashed line. The number counts are enhanced in this bin by a factor of $\sim 3$ compared to the surrounding redshift bins. {\bf Right:} Redshift histogram in the GOODS-S field. The bin located at $z=6.73\pm 0.03$ is marked with a vertical dashed line, and serves as a control for the histogram in GOODS-N.}
\label{fig:spike}
\end{figure*}

\begin{figure}
\centering
\includegraphics[width=1.0\linewidth]{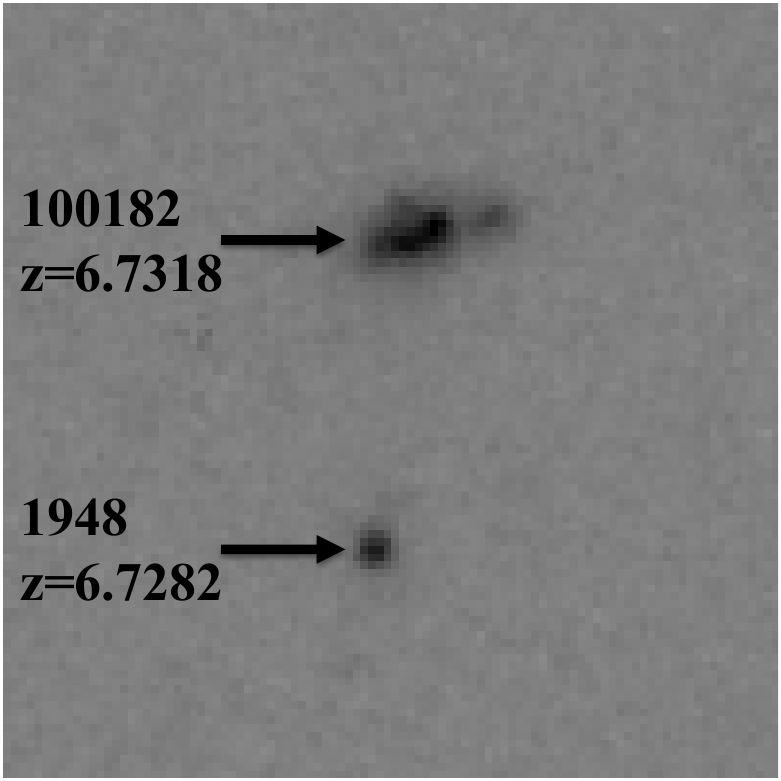}
\caption{Companion of GOODSN-100182. This 4"$\times$4" F356W postage stamp of GOODSN-100182 is oriented with North up and East to the left, and shows the spectroscopically-confirmed neighbor of GOODSN-100182. This galaxy is ID=1948 ($z_{\rm spec}=6.7282$) in the JADES Data Release 3 \citep{deugenio2024}, which is offset almost directly South of GOODSN-100182 by 1\secpoint6. This angular separation corresponds to 8.7 proper kpc in projected distance at the redshift of GOODSN-100182. The difference in redshift between GOODSN-1948 and GOODSN-100182 corresponds to $\Delta v=-140 \mbox{ km s}^{-1}$.
}
\label{fig:pair}
\end{figure}

\section{Discussion}
\label{sec:discussion}

\subsection{Environment of GOODSN-100182}
\label{sec:discussion-environment}
The maturity of GOODSN-100182 motivates an investigation of its larger-scale cosmic environment. It is a fairly basic prediction of standard galaxy formation models that galaxies within large-scale overdensities should be older than those outside \citep[e.g.,][]{behroozi2013}.
Recent observations have shown that massive, dusty $z>5$ galaxies are associated with redshift overdensities \citep[e.g.,][]{xiao2024,herarddemanche2024,arribas2024}. At the same time, galaxies within redshift overdensities at $z>5$ show evidence of accelerated formation in their best-fit stellar population parameters \citep[e.g.,][]{helton2024,morishita2024b}, building on earlier work demonstrating that $z\sim 2$ star-forming protocluster galaxies are older and more massive than their field counterparts \citep{steidel2005}.

We used the publicly available spectroscopic redshift catalog of the GOODS-N field from the JADES survey \citep{deugenio2024} to construct a redshift histogram in the GOODS-N field at $z=6-9$. As shown in Figure~\ref{fig:spike}, left, we find an apparent redshift spike at $z=6.73\pm 0.03$. We construct an analogous histogram for the redshifts collected in the GOODS-S field by JADES (Figure~\ref{fig:spike}, right). This histogram serves as a control since the JADES program used the same selection function in both fields. There is no corresponding $z=6.73$ redshift spike in GOODS-S, consistent with there being a $z=6.73$ ovedensity in GOODS-N.

On a more local scale, GOODSN-100182 has a companion galaxy at a projected separation of 1\secpoint63 to the south, with ID=1948 in the JADES data release of \citet{deugenio2024}. GOODSN-1948 is spectroscopically confirmed in the JADES catalog with $z=6.7282$. Accordingly, it is offset by 8.7 proper kpc in projected separation and $-140 \mbox{ km s}^{-1}$ in velocity space. As suggested by Figure~\ref{fig:pair}, in which GOODSN-1948 appears significantly more compact than GOODSN-100182 ($r_{\rm eff}=0.36\pm 0.03$~kpc, based on the average of F356W and F444W sizes), the properties of GOODSN-1948 are overall quite distinct from those of GOODSN-100182: the best-fit model of its SED yields a stellar mass of $\sim 10^9 M_{\odot}$; the H$\alpha$/H$\beta$ ratio in its NIRSpec spectrum suggests an absence of dust; and its additional emission-line properties (e.g., undetected [NII]$\lambda6583$, and an [OIII]$\lambda5007$/H$\beta$ ratio of 6) suggest a significantly less chemically-enriched ISM than that of GOODSN-100182. Given the small angular and velocity separations between GOODSN-1948 and GOODSN-100182, the two galaxies are likely gravitationally interacting. Given their stellar mass ratio, this galaxy pair then represents the early stages of a minor-merger event. While GOODSN-100182 is surrounded by $\sim 50$ additional galaxies within a radius of 10" ($\sim 50$ proper kpc in projection), we use the photometric redshift catalog available in the Dawn {\it JWST} Archive to show that none besides GOODSN-1948 has a redshift within $\Delta z=1$ of GOODSN-100182.

\subsection{Uniqueness of GOODSN-100182}
\label{sec:discussion-unique}
We have demonstrated many ways in which GOODSN-100182 stands apart from the bulk of the star-forming galaxy population at $z\sim 7$ \citep[e.g.,][]{topping2022,endsley2023,robertsborsani2024}. This galaxy has a significantly redder UV slope ($\beta$) and larger stellar mass  than the average $z\sim 7$ star-forming galaxy with the same rest-frame UV luminosity. Its rest-optical emission line spectrum also suggests a significant amount of dust attenuation, and a degree of metal enrichment in excess of even galaxies at $z\sim 2-3$ with similar stellar masses. In fact, in terms of [NII]$\lambda6583$/H$\alpha$ and [NII]$\lambda6583$/[OII]$\lambda3727$, GOODSN-100182 shares properties with typical $z\sim 0$ galaxies of the same stellar mass. Furthermore, accounting for the correlation between size and stellar mass, we find that the radius of GOODSN-100182 is significantly larger than the median observed among a sample of $10^{10} M_{\odot}$ $z\sim 7$ galaxies \citep{chworowsky2024}. The spatial extent of GOODSN-100182 is  much closer to the expectations for galaxies at $z\sim 3$. Considering in concert key scaling relations of galaxies such as the main sequence  and mass-metallicity relation, we find that GOODSN-100182 is elevated in {\it both} SFR and (indirectly inferred) oxgyen abundance relative to the average given its stellar mass. Such {\it correlated} deviations in SFR and metallicity at fixed mass  run counter to the observations and theoretical explanations of the so-called Fundamental Metallicity Relationship (FMR), in which deviations in SFR and metallicity at fixed mass are {\it anti-correlated} \citep[e.g.,][]{ellison2008,mannucci2010,laralopez2010,dave2017}.

In addition to standing out from blue, metal-poor star-forming galaxies at $z\sim 7$, GOODSN-100182 is also distinct from other {\it red} galaxy populations uncovered by recent {\it JWST} observations. For example, it does not satisfy the color criteria of so-called ``Little Red Dots" (LRDs) \citep[e.g.,][]{labbe2023,greene2024,kokorev2024}. Such sources are typically identified by a color of $F277W-F444W>1.0$ (or even $F277W-F444W>1.5$ \citealt{akins2024}), whereas GOODSN-100182 has $F277W-F444W=0.60\pm0.02$. In the original ``v-shaped" criteria proposed by \citet{labbe2023}, the rest-UV color requirement was $F150W-F277W<0.7$, which is also not satisfied by GOODSN-100182, given its $F150W-F277W$ color of $1.12\pm0.05$. Therefore, GOODSN-100182 is both too blue in rest-optical color, and too red in rest-UV color, to be identified as an LRD. Given its large physical extent, GOODSN-100182 is also structurally distinct from LRDs.

Another red-galaxy selection aims to find ``HST-Dark" galaxies missing from previous censuses of the $z>3$ galaxy population, using $F160W-F444W>2$ and the $H$-band magnitude ($F160W$ or $F150W$) fainter than a given limit (i.e., 25, or 27 AB) \citep[e.g.,][]{barrufet2023,williams2024,gottumukkala2024}. Physically, such empirical criteria yield sources with $A_V>2$.
GOODSN-100182 shows evidence for significant dust attenuation, yet its rest-frame UV/optical colors and inferred $A_V$ are not quite so extreme, with $F150W-F444W=1.72\pm0.05$ and $A_V=1.6$. Furthermore, GOODSN-100182 was not ``HST-Dark", having actually been identified in {\it HST} surveys \citep{bouwens2015,jung2020}. While the photometric properties of GOODSN-100182 could be discerned using pre-{\it JWST} facilities, its large rest-optical size and unusual emission-line spectrum were entirely beyond reach until now.

Finally, the REBELS survey, an ALMA large program targeting UV-luminous galaxies at $z>6.5$, also confirmed the existence of massive, dusty galaxies at $z\sim 7$ with both [CII]$\lambda 158 \mu$m and dust continuum detections \citep{bouwens2022, inami2022}. Although GOODSN-100182 overlaps some of the parameter space spanned by the REBELS sample, it also shows some key complementarities. The full REBELS sample contains 49 galaxies, all with $M_{\rm UV}\leq -21.3$.  \citet{bowler2024} divide the 20 REBELS galaxies at $z=6.5-6.9$ into two bins of $M_{\rm UV}$, with average values of 
$\langle M_{\rm UV}\rangle=-22.34$ and $-21.71$, respectively, for the brighter and fainter bins. The corresponding average UV slopes, $\beta$ are, respectively, $\langle \beta \rangle = -2.10$ and $-1.79$. On average, therefore, REBELS $z\sim 7$ galaxies are both significantly brighter and bluer in the rest-UV than GOODSN-100182, and no individual galaxy in REBELS is as faint or red in the rest-UV as GOODSN-100182. The average stellar masses ($\log(M_*/M_{\odot})=9.5$) and SFRs ($40 M_{\odot}{\rm yr}^{-1}$) for these $z\sim 7$ REBELS galaxies are also lower than that of GOODSN-100182, however, the high-mass / high-SFR tail of the REBELS distribution overlaps with GOODSN-100182. In particular, the galaxy REBELS-25 ($z=7.31$) has  comparable stellar mass ($\log(M_*/M_{\odot})=9.9$) and even higher dust-corrected SFR ($199 M_{\odot}{\rm yr}^{-1}$), based the sum of UV and dust-continuum emission \citep{rowland2024}. In addition, while the rest-UV morphology of REBELS-25 inferred from {\it HST}/WFC3 imaging shows multiple compact clumps, the [CII] map of REBELS-25 reveals an extended ($r=2$~kpc), rotating disk. Therefore, both GOODSN-100182 and REBELS-25 show evidence for large stellar mass, SFR, and physical extent. The key differences are in rest-UV luminosity and color ($M_{UV}=-21.67$ and $\beta=-1.85$ for REBELS-25, whereas $M_{UV}=-20.28$ and $\beta=-0.50$ for GOODSN-100182). The rest-optical spectrum, and therefore ISM excitation and metallicity, of REBELS-25 is still unknown,
just as the [CII] and dust continuum properties of GOODSN-100182 currently are. However, based on the data available thus far, both galaxies provide evidence for the existence of mature systems at $z\sim 7$.

\subsection{Is GOODSN-100182 an AGN?}
\label{sec:discussion-agn}
A couple of features of GOODSN-100182 raise the question of whether an AGN is present. These include the broad emission-line component in H$\alpha$ (Section~\ref{sec:results-spectrum}) and the unresolved (point-source) component that is more prominent at bluer wavelengths (Section~\ref{sec:results-props-structure}, Figure~\ref{fig:stamps-100182}). 

In particular, one key question is whether the broad H$\alpha$ emission is itself associated with the point-source component. We note here first that the NIRSpec slit for GOODSN-100182 only partially covers the unresolved component. Second, the rest-optical spectrum of GOODSN-100182 falls within the observed wavelength range of the F444W filter, in which the point-source represents $<10$\% of the galaxy light distribution. In fact, guided by the light-weighted centroid of the continuum, the extracted spectrum of GOODSN-100182 in the G140M grating (in which the point source is significantly more prominent, and in which we detect no significant emission lines) is spatially offset by 0\secpoint15 to the southwest along the slit from the spectra in G235M and G395M (in which light from other regions of the galaxy make a larger contribution, and in which all of the strong rest-optical emission lines are detected). This offset suggests that the rest-optical emission lines within the G395M grating (H$\beta$, [OIII]$\lambda5007$, H$\alpha$, [NII]$\lambda6583$) are not dominated by emission from a point source. Furthermore, even at shorter wavelengths, there are no significant detections of C~IV$\lambda\lambda1548,1550$, Mg~II$\lambda\lambda2796,2803$, or [NeV]$\lambda3426$ emission, features that are commonly associated with the presence of an AGN.

As for the kinematics of H$\alpha$ emission, with a single unresolved slit spectrum only partially sampling the extended light distribution of GOODSN-100182, there is insufficient information to determine the underlying scenario. This could include an AGN, an outflow, or simply the complex kinematics of an extended disk \citep{nelson2024}. In order to disentangle the different kinematic components of H$\alpha$ and their origins, we require a spatially-resolved map of the emission of GOODSN-100182, covering both the point source and the surrounding emission with many distinct spatial elements. The NIRSpec integral field unit (IFU) capability provides the ideal set-up for such observations.

\subsection{Comparisons With Theory}
\label{sec:discussion-theory}
The overall narrative of the evolution of the galaxy population back to the earliest times focuses on the average: more compact morphologies, bluer colors, lower metallicities. In addition to understanding the average galaxy properties at the highest redshifts, we also must attempt to explain the extremes of the galaxy population. {\it JWST} has revealed one such extreme in GOODSN-100182. While the mass of this galaxy may not strain our $\Lambda$CDM cosmological framework, its size, dust content, metallicity, and SFR do test the baryonic physics of early galaxy assembly. For example, the emission line and structural measurements of GOODSN-100182 provide new constraints on simulations such as the Cosmic Sands suite presented in \citet{lower2023}, who find that comparably massive galaxies at $z\sim 7$ have similar SFRs to that of GOODSN-100182, and sometimes disk-like morphologies, but smaller sizes ($r_{\rm eff}<1$~kpc). Other important constraints on models of massive galaxy formation will come from an estimate of the dust mass in GOODSN-100182 -- not simply the significant apparent dust attenuation -- which will be possible based on millimeter-wave, dust-continuum observations. A comparison of the dust to stellar mass ratio  with those predicted for $z\sim 7$ galaxies in cosmological simulations \citep[e.g.,][]{dicesare2023,lower2023,choban2024} will provide insights into the processes driving the early formation and destruction of dust grains in galaxies.

\section{Summary}
\label{sec:summary}
We have presented the remarkable properties of GOODSN-100182, a massive, $z=6.73$ star-forming galaxy observed as part of the AURORA {\it JWST}/NIRSpec survey. The combined imaging and spectroscopic dataset reveals GOODSN-100182 to stand apart from all other $z\sim 7$ star-forming galaxies surveyed to date. In particular, we find:

\begin{enumerate}
    \item GOODSN-100182 lies at the extreme of rest-UV-selected $z\sim 7$ star-forming galaxies in terms of its photometric properties. Given its absolute UV luminosity of $M_{\rm UV}=-20.3$, GOODSN-100182's rest-UV slope of $\beta=-0.5$ and stellar mass of $\sim 10^{10} M_{\odot}$ are significantly redder and more massive than average. 

    \item GOODSN-100182 is also anomalous in terms of its structural properties. It is significantly larger in size than other galaxies at either comparable UV luminosities or stellar masses, featuring a prominent disk-like component revealed for the first time at rest-optical wavelengths by {\it JWST}/NIRCam, with $r_{\rm eff}\sim 1.5$~kpc and S\'ersic index $n\sim 1$. Although this galaxy was detected at $\lambda >3 \mu$m with {\it Spitzer}/IRAC prior to the launch of {\it JWST}, the higher-angular-resolution capabilities of NIRCam enable a characterization of the size and shape of the galaxy for the first time.

    \item The spectrum of GOODSN-100182 features strong [NII]$\lambda6583$ emission and relatively low excitation and ionization compared to the bulk of $z\sim 7$ star-forming galaxy spectra measured  to date with {\it JWST} spectroscopy. Given its stellar mass of $10^{10}M_{\odot}$, the degree of dust attenuation and chemical enrichment in GOODSN-100182 is most consistent with that of a galaxy at $z\leq 2$ rather than $z\sim 7$. The spectrum of GOODSN-100182 also reveals evidence for a broad kinematic component in H$\alpha$ emission. The full interpretation of this complex emission-line profile will require spatially-resolved IFU spectroscopy. 

    \item GOODSN-100182 resides within a large-scale galaxy overdensity and, furthermore, has a close companion at a projected separation of 8.7 proper kpc and $-140 \mbox{ km s}^{-1}$ in velocity space. The galaxy overdensity surrounding GOODSN-100182 may represent a site of accelerated galaxy formation in the early universe.

\end{enumerate}

A more systematic search for massive ($M_*>10^{10}M_{\odot}$) systems that are also large in spatial extent ($r_{\rm eff} > 1$~kpc) must now be conducted to quantify the space density and properties of such extreme objects at the earliest times. The properties of such galaxies will place unique and powerful constraints on models of galaxy formation.

\section*{Acknowledgements}
This work is based on observations made with the NASA/ESA/CSA James Webb Space Telescope. The data were
obtained from the Mikulski Archive for Space Telescopes at
the Space Telescope Science Institute, which is operated by the
Association of Universities for Research in Astronomy, Inc.,
under NASA contract NAS5-03127 for JWST.  The specific observations analyzed can be accessed via \dataset[DOI: 10.17909/hvne-7139]{https://archive.stsci.edu/doi/resolve/resolve.html?doi=10.17909/hvne-7139}.
We also acknowledge support from NASA grant JWST-GO-01914. FC acknowledges support from a UKRI Frontier Research Guarantee Grant (PI Cullen; grant reference: EP/X021025/1).
ACC thanks the Leverhulme Trust for their support via a Leverhulme Early Career Fellowship.  CTD, DJM, RJM, and JSD acknowledge the support of the Science and Technology Facilities Council. JSD also acknowledges the support of the Royal Society through a Royal Society Research Professorship.
RD acknowledges support from the Wolfson Research Merit Award program of the U.K. Royal Society.  
KG acknowledges support from the Australian Research Council Laureate Fellowship FL180100060.
MK acknowledges funding from the Dutch Research Council (NWO) through the award of the Vici grant VI.C.222.047 (project 2010007169).
PO acknowledges the Swiss State Secretariat for Education, Research and Innovation (SERI) under contract number MB22.00072, as well as from the Swiss National Science Foundation (SNSF) through project grant 200020$\_$207349.
AJP was generously supported by a Carnegie Fellowship through the Carnegie Observatories.
DN was funded by JWST-AR-01883.001.

\bibliographystyle{apj}
\bibliography{aurora-z7}

\end{document}